\documentclass[twocolumn,showpacs,preprintnumbers]{revtex4}

\usepackage{epsfig}

\begin{document}

\preprint{}

\title{The polymer phase of the TDAE-C$_{60}$ organic ferromagnet}
\author{Slaven Garaj$^1$, Takashi Kambe$^{1,2}$, L\'aszl\'o Forr\'o$^1$, Andrzej Sienkiewicz$^3$,\\ Motoyasu Fujiwara$^2$, Kokichi
Oshima$^2$}

\affiliation{$^1$ Institute of Physics of Complex Matter, \'Ecole
Polytechnique F\'ed\'erale de Lausanne, CH-1015 Lausanne,
Switzerland}

\affiliation{$^2$ Graduate school of natural science and
technology, Okayama University, 3-1-1 Tsushima, Okayama 700-8530,
Japan}

\affiliation{$^3$ Institute of Physics, Polish Academy of
Sciences, 02-668 Warsaw, Poland}

\date{\today}

\begin{abstract}
The high-pressure Electron Spin Resonance (ESR) measurements were
preformed on TDAE-C$_{60}$ single crystals and stability of the
polymeric phase was established in the $P - T$ parameter space. At
7 kbar the system undergoes a ferromagnetic to paramagnetic phase
transition due to the pressure-induced polymerization. The
polymeric phase remains stable after the pressure release. The
depolymerization of the pressure-induced phase was observed at
the temperature of 520 K. Below room temperature, the polymeric
phase behaves as a simple Curie-type insulator with one unpaired
electron spin per chemical formula. The TDAE$^+$ donor-related
unpaired electron spins, formerly ESR-silent, become active above
the temperature of 320 K and the Curie-Weiss behavior is
re-established.
\end{abstract}

\maketitle

\section{Introduction}
The organic charge-transfer compound TDAE-C$_{60}$ (where TDAE is
tetrakis-dimethylamino-ethylene) is a ferromagnet with the Curie
transition temperature of $T_C=$ 16 K (Ref. \cite{1}). This is the
highest temperature onset of ferromagnetic behavior for a purely
organic material. It is customarily accepted that the system is an
isotropic Heisenberg ferromagnet with an extremely small
anisotropy field\cite{2}\cite{3}.

The TDAE molecule, a strong electron donor, transfers one
electron to the lowest unoccupied molecular orbital (LUMO) of
C$_{60}$ in a similar way to that found in alkali metal C$_{60}$
compounds.  It has been established that the single-charged
C$_{60}$ alkali salts, A$_{1}$C$_{60}$ (A= K, Rb, Cs), reveal
metallic properties in a wide temperature range\cite{4}. Although,
analogously to the alkali metal C$_{60}$ compounds, the valence
band in TDEA-C$_{60}$ originates from the triply degenerated
$t_{1g}$ orbital of the C$_{60}$ molecule, the TDAE-C$_{60}$
system was found to be non-metallic\cite{5}\cite{6}. The
non-metallic behavior of TDAE-C$_{60}$ was explained assuming the
Jahn-Teller effect and Mott-Hubbard localization\cite{7}\cite{8}.

Magnetic susceptibility measurements\cite{9}\cite{10}\cite{11} as
well as Electron Spin Resonance (ESR) results\cite{11}\cite{12}
showed that the TDAE-C$_{60}$ system has only one $S=1/2$ magnetic
moment per chemical formula unit. From the $g$-factor analysis it
was concluded that the spins are mainly localized on C$_{60}^-$
(Ref. \cite{11}). The TDAE$^+$ donors should also carry
non-paired electrons. However, these centers remain ESR silent,
which is probably due to the mechanism of the spin-singlet pairing
resulting from the dimerization shift of the neighboring TDAE$^+$
molecules\cite{13}.

Two magnetic variants of TDAE-C$_{60}$ were discovered. As grown
crystals ($\alpha'$-phase) reveal the antiferromagnetic (AFM)
behavior, whereas annealed crystals ($\alpha$-phase) show the
transition to the ferromagnetic (FM) state\cite{14}.
Surprisingly, in X-ray diffraction (XRD), these two structures
look almost indistinguishable around the room temperature when
the C$_{60}$ balls rotate freely. However, the difference is
observed when freezing of the rotation of the C$_{60}$ balls
occurs at low temperatures\cite{12}\cite{15}. In the FM phase, a
new Bragg reflection develops, suggesting lowering of symmetry
and ordering of the C$_{60}$ balls\cite{12}. This reflection is
not observed in the AFM phase, which suggests the existence of a
disordered (glassy) state of the C$_{60}$ orientations.

Nuclear Magnetic Resonance (NMR) data imply that in the FM phase
the electron spin of C$_{60}^-$ are partially delocalized on the
TDAE molecule, thus raising a possibility for the super-exchange
interaction\cite{16}. In a simple model, one would expect that
the super-exchange mechanism would promote the AFM behavior.
Nevertheless, the observed ferromagnetic properties are explained
in terms of the Hubbard model with the antiferro-orbital order of
the Jahn-Teller distorted (JTD) C$_{60}$ balls\cite{8}. Indeed,
NMR observations of the JT distortion and the asymmetric charge
distribution on the C$_{60}$ balls have recently been
reported\cite{16a}. However, the NMR results of Arcon et
al.\cite{16} suggest that the $\alpha$-phase contains the
C$_{60}^-$ molecules with both ferromagnetic and
antiferromagnetic configurations. In that case, the spontaneous
magnetization should appear when the concentration of the FM
configuration is high enough for the appearance of an infinite
ferromagnetic cluster through percolation
mechanism\cite{15}\cite{16b}.

Recently, Mizoguchi and co-workers\cite{17} reported
polymerization of TDAE-C$_{60}$ under pressure of $\sim 10$ kbar.
The polymerized phase ($\beta$-phase) remains stable even after
releasing the pressure. The polymerization process occurs along
the c-axis and it has been suggested that the linear polymers can
be formed due to a [2+2] cycloaddition process.

In this article, we report the $P - T$ diagram of the stability of
the polymeric TDAE-C$_{60}$ structure. The effects of the
high-hydrostatic pressure on the physical properties of both
monomeric and polymeric phases were investigated. We studied the
physical properties of the $\beta$-phase at low and high
temperatures and the effect of the temperature-induced
depolymerization. The properties of the TDAE-C$_{60}$ polymer are
compared with those of other bonded fullerene structures.

\begin{figure}[tbp]
\centerline{ \epsfxsize=0.45 \textwidth{\epsfbox{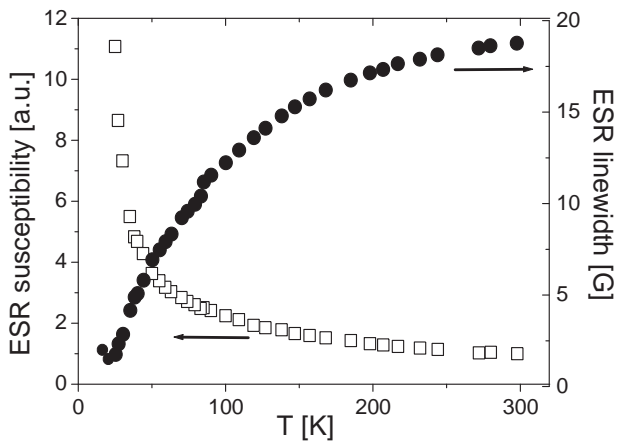}} }
\caption{Spin susceptibility (left scale, open squares) and the
ESR linewidth (right scale, filled circles) of the ferromagnetic
TDAE-C$_{60}$ single crystal as a function of temperature. In the
ferromagnetic region below $T_C= 16$ K, the linewidth cannot be
easily defined because the line shapes are strongly disported due
to mosaicity of the crystal} \label{Figure1}
\end{figure}

\section{Experiment}

Single crystals of the TDAE-C$_{60}$ were prepared by the
diffusion method as reported in Ref. \cite{12}. The single
crystals dimensions were, typically, of $0.3 \times 0.3\times
0.3$ mm$^3$. The presence of the FM phase was checked by the
magnetization measurements.

\begin{figure}[tbp]
\centerline{ \epsfxsize=0.40 \textwidth{\epsfbox{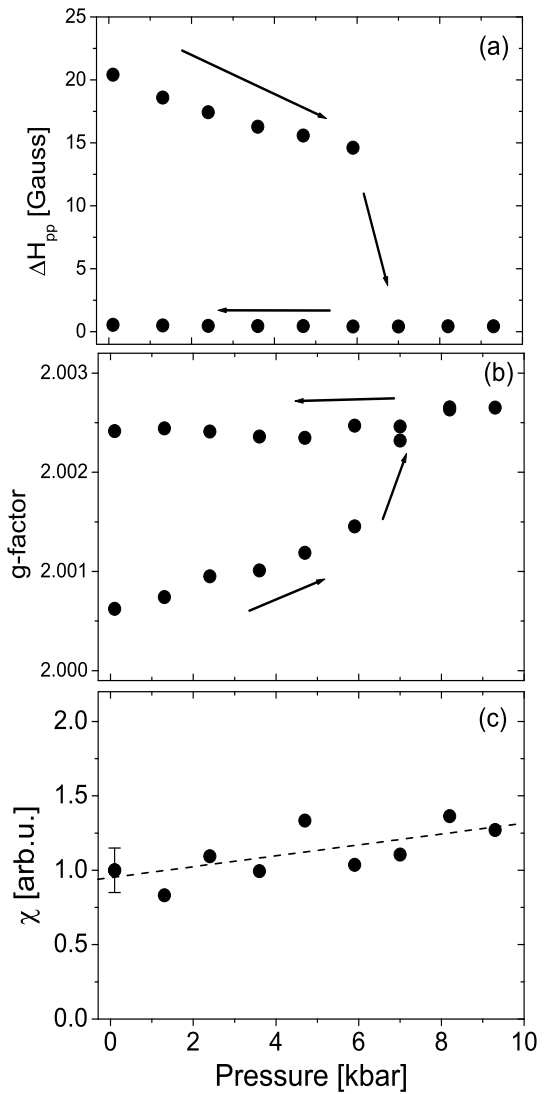}} }
\caption{ Pressure dependence of the ESR parameters for single
crystal TDAE-C$_{60}$ at room temperature: (a) linewidth; (b)
$g$-factor; (c) relative spin susceptibility. Phase transition to
the polymerized phase is visible at $P_C= 7$ kbar. }
\label{Figure2}
\end{figure}

Ambient pressure Electron Spin Resonance (ESR) measurements were
performed in the temperature range of $5 - 600$ K using a Bruker
ESP300E X-band spectrometer.  In the temperature range of $5 -
300$ K the ESR spectra were acquired using a standard Bruker
TE$_{102}$ cavity that was equipped with an Oxford Instrument,
Model ESR900, gas-flow cooling system. In the upper temperature
range ($300 - 600$ K) we used a Bruker ER4114HT high-temperature
cavity system.  The magnetic field and the microwave frequency
were calibrated using a commercially available NMR gauss-meter
and a frequency counter, respectively. The ESR line intensities
were calibrated using a secondary standard sample, a small speck
of DPPH (2,2-dipenyl-1-picrylhydrazyl from Sigma).

The high-pressure ESR measurements were performed using a
high-pressure system that was recently developed at the EPFL. The
high-pressure ESR probe was designed as an interface to our
Bruker ESP300E X-band spectrometer.  The probe consists of two
sub-assemblies: 1) the microwave resonant structure containing
the double-stacked Dielectric Resonator (DR) and 2) the miniature
sapphire-anvil pressure cell (SAC).  The SAC is ruby-calibrated;
thus the hydrostatic pressure can be monitored in situ by
detecting the pressure-induced shift of the red fluorescence of
Cr$^{3+}$ ions in a small crystal of ruby. The commercially
available "Daphne" oil was used as a pressure-transmitting
medium.  In this work, the maximum applied pressure was of 9
kbar. For performing low temperature measurements, the
high-pressure probe was inserted into the CF-1200 Oxford
Instrument gas-flow cryostat operating in the $5 - 290$ K
temperature range. The details of the DR-based high-pressure ESR
probe will be published shortly elsewhere \cite{18}. The ESR line
intensities and the $g$-factor were calibrated using an
additional reference sample, a polycrystalline MnO/MgO, which was
positioned in the active zone of the microwave resonant structure
(close to the gasket of the SAC).

\begin{figure*}[tbp]
\centerline{ \epsfxsize=0.85 \textwidth{\epsfbox{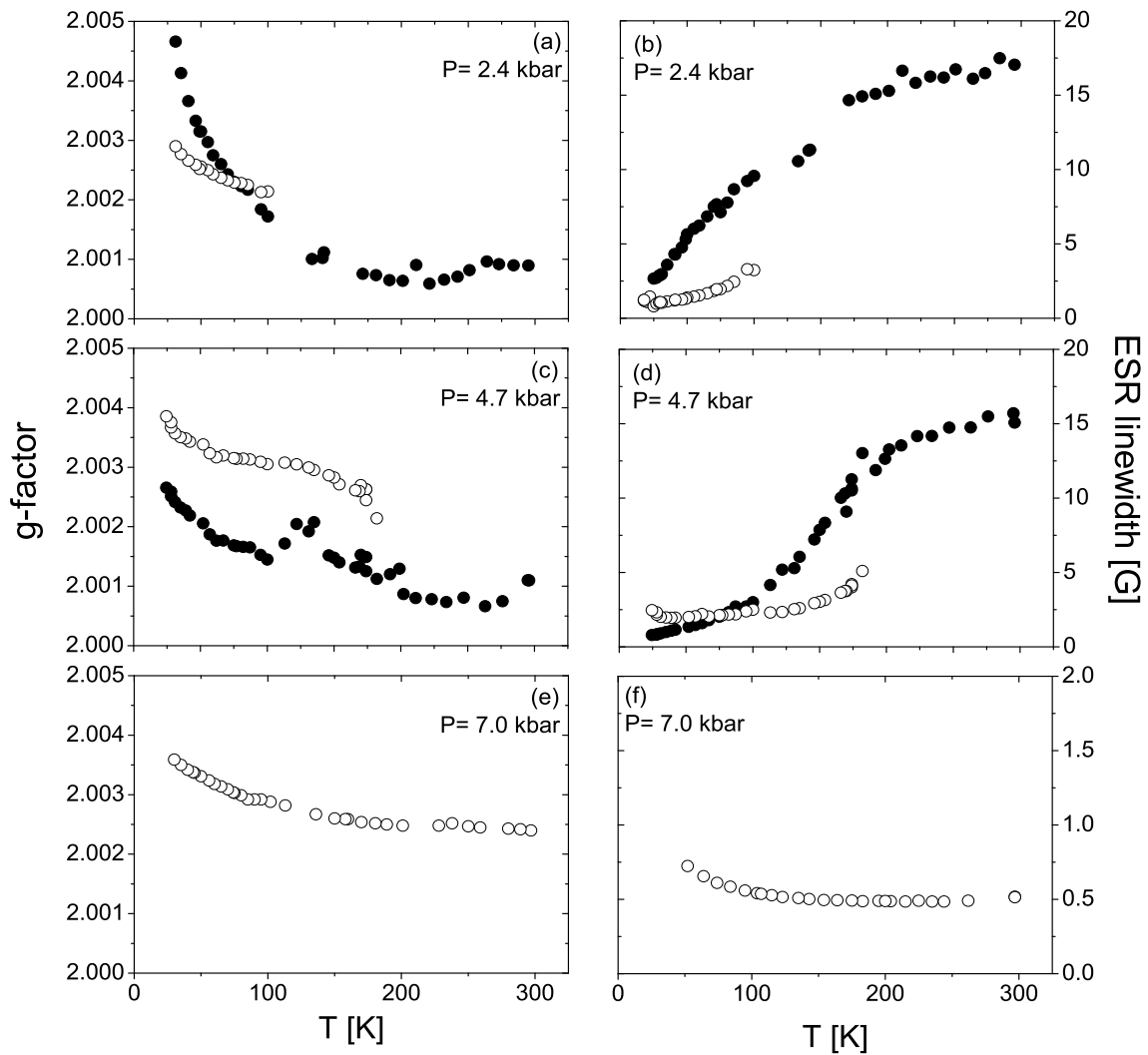}} }
\caption{ Temperature dependence of the ESR linewidth and the
$g$-factor for the pressures of 2.6 kbar (a, b), 4.7 kbar (c, d,)
and 7 kbar (e, f). Below the polymerization temperature, in
addition to monomeric line (filled circles), a new line appears
(opem circles) which was assigned to polymer. At 7 kbar the system
is fully polymerized at the room temperature. } \label{Figure3}
\end{figure*}

\section{Results}

\subsection{Pressure-induced polymerization}

Figure~\ref{Figure1} shows the temperature evolution of the
ESR-probed spin susceptibility and the ESR linewidth of the
ferromagnetic phase of TDAE-C$_{60}$ measured at ambient
pressure.  In the ferromagnetic region, below $T_C= 16$ K, the
linewidth cannot be determined precisely due to the strong
distortion of the ESR line shape. The line shape distortion is
probably due to the non-homogeneity of the local internal fields
in different ferromagnetic domains of the crystal.

The pressure dependence of the ESR linewidth (peak-to-peak,
$\Delta H_{pp}$) of TDAE-C$_{60}$ at ambient temperature is shown
in Fig.~\ref{Figure2}a, whereas Fig.~\ref{Figure2}b shows the
pressure dependence of the $g$-factor. The initial (ambient
pressure) linewidth of $\Delta H_{pp} \approx 20$ G slowly
decreases with increasing pressure to 15 G at $P= 6$ kbar. Then,
at $P_C = 7$ kbar, a transition to the polymeric phase is clearly
seen as a sudden drop in the ESR linewidth (Fig.~\ref{Figure2}a).
This is also accompanied by an abrupt change (increase) in the
$g$-factor (Fig.~\ref{Figure2}b). The phase transition is
irreversible and the polymeric phase remains stable after
releasing the pressure. These results are in a partial agreement
with the ESR measurements performed at lower microwave
frequencies by Mizoguchi et al.\cite{17}. The major difference
is, however, that the critical pressure of suppression of the
ferromagnetic transition in our measurements is rather lower and
coincides with the pressure of the complete polymerization.

In the monomeric TDAE-C$_{60}$ the ESR linewidth is defined by
dipolar interaction with additional narrowing introduced by the
exchange interaction. Accordingly, as can be seen in
Fig.~\ref{Figure2}a, the monomer's linewidth slowly narrows with
pressure approaching 7 kbar. This is due to the larger overlap of
the electronic wave functions, which leads to an enhanced
exchange interaction. At ambient pressure, the $g$-factor value
for the monomeric phase of TDAE-C$_{60}$ is 2.0006 and is
distinctively closer to the $g$-factor of the fullerene anion
C$_{60}^-$ ($g_{C_{60}^-} = 1.9998$) than to the $g$-factor value
that is characteristic for the TDAE$^+$ cation ($g_{TDAE^+} =
2.0036$). This is due to the fact that the ESR signal originates
from the electrons that are mainly localized on the C$_{60}$
balls.  The spins on the TDAE$^+$ radicals are ESR-silent, which
is probably due to a slight dimerization of TDAE molecules, thus
yielding a spin-singlet configuration. With increasing pressure,
the $g$-factor linearly increases towards the value of the
TDAE$^+$ cation. This implies that the unpaired spin density is
spreading towards the TDAE molecule with increasing pressure.

At $P_C=$ 7 kbar, a sudden narrowing of the ESR line is visible
due to polymerization. The linewidth drops by two orders of
magnitude, reaching $\Delta H_{pp}$(polymer)= 0.5 G. The narrow
linewidth and the Currie type temperature dependence of the spin
susceptibility at this pressure suggest that the polymeric phase
is non-metallic. Upon polymerization the $g$-factor value rises
reaching 2.0024 and becomes almost pressure independent.

Mizoguchi et al.\cite{17} suggested that the polymerization
process decouples the previously ESR-silent spins related to the
TDAE$^+$, which should lead to an effective doubling of the total
number of spins.  The observed pressure dependence of ESR
susceptibility at room temperature (Fig.~\ref{Figure2}c) does not
support this suggestion. At this stage, a small increase (of
$\sim 20\%$) of the ESR susceptibility upon polymerization can be
explained by the differences in the Weiss temperatures between
the two phases.

\subsection{Coexistence of phases}

Figure~\ref{Figure3} shows the temperature dependence of the ESR
linewidth and the $g$-factor for three different pressures. The
temperature dependence of the linewidth and the $g$-factor at
applied pressure of $P=$ 2.6 kbar is shown in Fig.~\ref{Figure3}a
and Fig.~\ref{Figure3}b, respectively.  As can be seen in
Fig.~\ref{Figure3}a, below the characteristic temperature, $T_P=
100$ K, in addition to the ESR line of TDAE-C$_{60}$ monomer
(full circles), a new line appears (open circles).  This new ESR
line can be assigned to the polymeric phase while taking into
account the $g$-factor and ESR linewidth evolution with pressure.
The intensity of the polymeric line does not exceed 15\% of the
intensity of the monomeric line. The temperature dependence of
the $g$-factor and ESR linewidth at $P=$ 4.7 kbar is shown in
Fig.~\ref{Figure3}c and Fig.~\ref{Figure3}d, respectively.  At
this pressure, a partial polymerization starts at higher
temperatures, around $T_P \approx 180$ K.  The intensity of the
polymeric line is now much more pronounced. For both applied
pressures, the monomeric phase undergoes the ferromagnetic phase
transition, whereas the polymeric phase does not. At pressure of
$P=$ 7.0 kbar, the sample is fully polymerized already at room
temperature (Fig.~\ref{Figure3}e and Fig.\ref{Figure3}f). No
ferromagnetic phase transition is visible down to 5 K.  At this
applied pressure, the ESR susceptibility follows a simple Curie
law, without a detectable Weiss constant. The ESR linewidth is
very narrow ($0.5$ G) and is almost independent of temperature,
whereas the $g$-factor slightly changes with temperature.

\begin{figure}[tbp]
\centerline{ \epsfxsize=0.45 \textwidth{\epsfbox{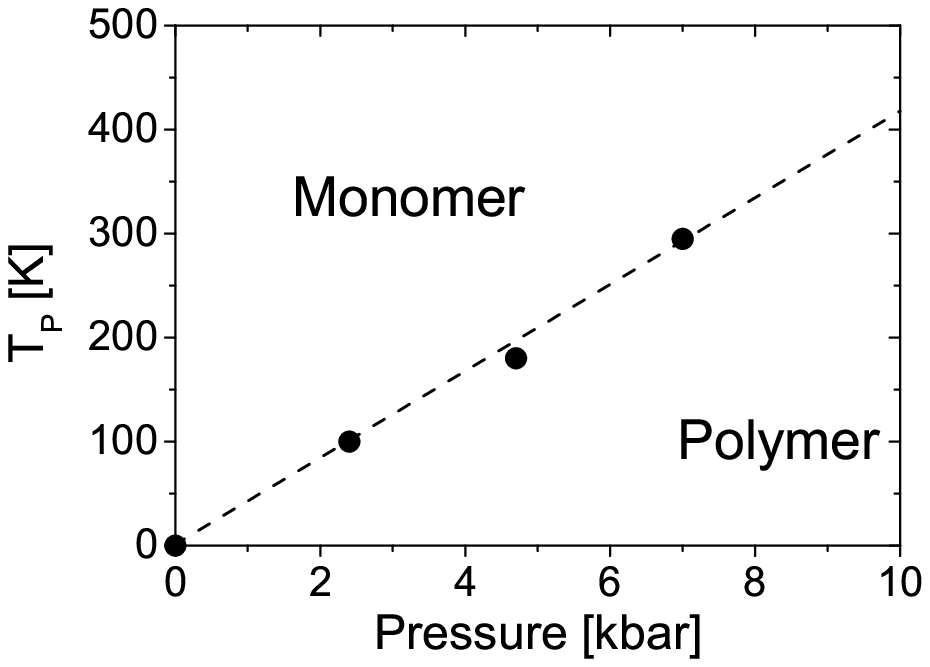}} }
\caption{ Polymerization temperature ($T_P$) as a function of
applied pressure. The polymerization temperature has a linear
dependence on the applied pressure, with the constant of
proportionality of  $\frac{dT_P}{P}=41 \pm 2$ K/kbar}
\label{Figure4}
\end{figure}

The stability of the polymer phase in the $P - T$ parameter space
is depicted in Fig.~\ref{Figure4}.  The polymerization
temperature ($T_P$) has a linear dependence with applied
pressure, where the proportionality constant. The ratio of the
polymeric and monomeric ESR line intensities depends not only on
pressure, but also on temperature. To deduce the exact structural
dynamics of the polymer formation an additional structural study
is needed.

\begin{figure}[tbp]
\centerline{ \epsfxsize=0.40 \textwidth{\epsfbox{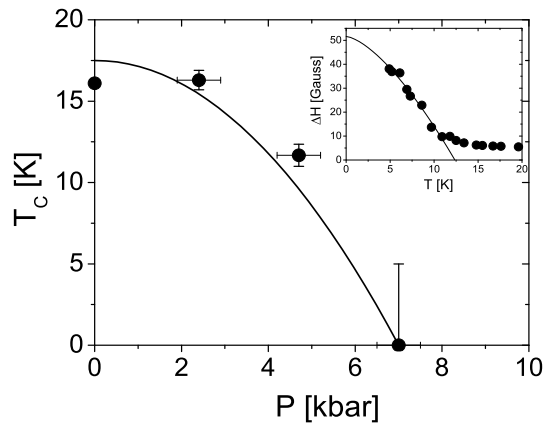}} }
\caption{ Pressure dependence of the ferromagnetic transition
temperature, $T_C$, for single crystal TDAE-C$_{60}$. $T_C$ at
each pressure is determined by the onset temperature of the
broadening of the linewidth distribution of the monomeric signal.
The lines are guides for the eye. Inset: line distribution
broadening for the pressure of 4.6 kbar.} \label{Figure5}
\end{figure}

The pressure dependence of the ferromagnetic transition
temperature ($T_C$) of the monomeric phase is depicted in
Fig.~\ref{Figure5}. To determine $T_C$, we cannot compare the
resonance field shift with the conventional Bloch's law, because
the resonance field has pronounced temperature dependence even
above $T_C$ (see Fig.~\ref{Figure3}). This shift is due to the
demagnetization effects, which depend on the sample
shape\cite{19}. Concomitantly, we define $T_C$ as the onset
temperature of the broadening of a linewidth distribution of the
monomer-related ESR features (see: inset to Fig.~\ref{Figure5}).
This linewidth distribution broadening is due to the growth of an
internal field below $T_C$ and the mosaicity of the crystal. The
observed pressure dependence of $T_C$ is similar to the parabolic
dependence reported in Ref. \cite{17}.  However, in contrast to
their ESR data acquired at low microwave frequencies and fields,
the ferromagnetic transition observed in this work vanishes
simultaneously with polymerization. This result can be expected
and explained in the framework of the theory by Kawamoto and
coworkers\cite{8}, because the onset of polymerization prevents
the antiferro ordering of the orbital of the JTD fullerene
molecules.

\subsection{Polymer phase at ambient pressure}

\begin{figure}[tbp]
\centerline{ \epsfxsize=0.43 \textwidth{\epsfbox{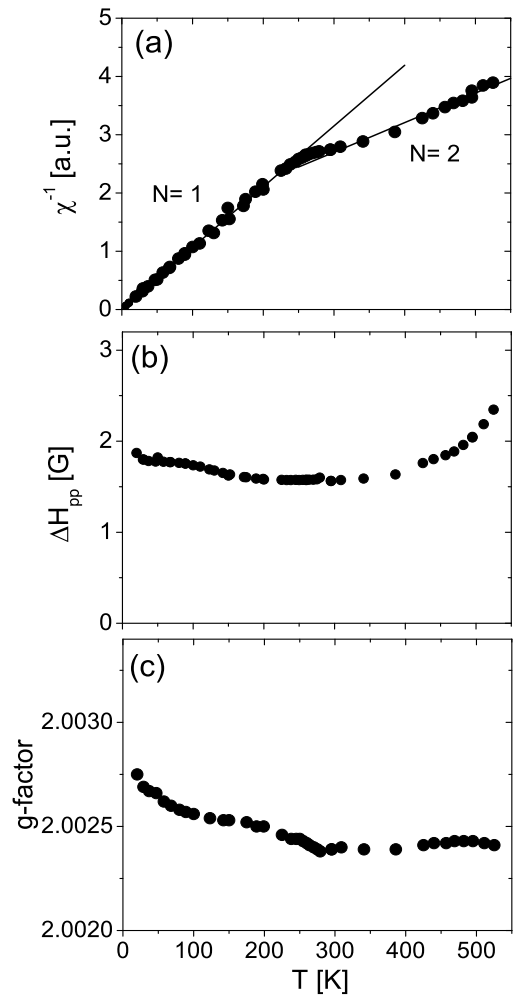}} }
\caption{  Temperature dependence of ESR parameters for polymeric
$\beta-$TDAE-C$_{60}$ phase at ambient pressures: (a) inverse
susceptibility (b) linewidth; (c) $g$-factor. In susceptibility,
the onset from simple Currie law is seen above 250 K, but the
full appearance of previously silent TDAE$^+$ spins is seen at
320 K, leading to doubling of number of spins and appearance of
the Weiss constant. The effect is probably connected to structural
dynamics of the TDAE dopant. } \label{Figure6}
\end{figure}

The polymeric phase remains stable even after the pressure is
released. For the polymeric phase, the temperature dependences of
the ESR parameters are shown in Fig.~\ref{Figure6}. As can be
seen from comparison of the results presented in
Fig.\ref{Figure6} and in Fig.~\ref{Figure3}a and
Fig.~\ref{Figure3}b, the low-temperature properties of the
polymeric phase at ambient pressure are very similar to those
observed at $P=$ 7 kbar.  The temperature dependence of the
inverse ESR susceptibility, $\chi^{-1}$, of the polymeric phase
is shown in Fig.~\ref{Figure6}a. Since we are dealing with
localized spins, this type of plot ($\chi^{-1}$ vs. T) is the most
informative, directly yielding information on the Curie constant
and the Weiss temperature of the system. The spin susceptibility
below the room temperature reveals one spin $S=$ 1/2 per chemical
formula unit ($N=$ 1) and a simple Curie behavior.

As seen in Fig.~\ref{Figure6}a, the inverse susceptibility
departs from the simple Curie behavior at 260 K. This transition
region is relatively broad and extends up to ca. 320 K. Above 320
K, the ESR susceptibility reveals the doubling of the number of
spins per chemical formula.  This phenomenon can be understood in
terms of re-appearance of the previously hidden spins of the
TDAE$^+$ radicals. The Weiss temperature does not vanish anymore,
having the value of $\Theta= 230 \pm 20$ K. This change of
behavior at high temperatures is also seen in the temperature
dependence of the linewidth (Fig.~\ref{Figure6}b). At lower
temperatures the $\Delta H_{pp}$ is almost constant, with a
slight tendency to decrease with increasing temperature, whereas
above the room temperature it changes its slope and starts to
increase more rapidly. The temperature dependence of the
$g$-factor (Fig.~\ref{Figure6}c) reveals a similar, distinctive
change of behavior above the room temperature.

\begin{figure}[tbp]
\centerline{ \epsfxsize=0.40 \textwidth{\epsfbox{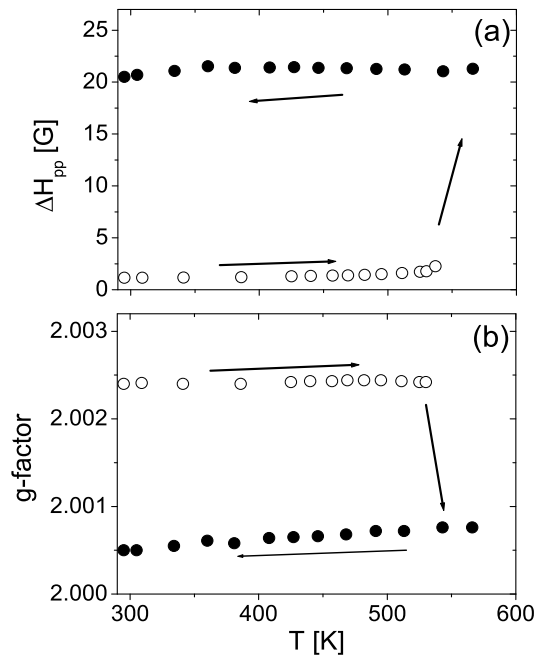}} }
\caption{Temperature dependence of the linewidth and the
$g$-factor of the polymerized crystal in the high-temperature
region. The polymeric phase is marked with open circles and the
monomeric with closed circles. The full depolymerization is
observed above 520 K. At the transition temperature, signals of
the both phases are detected, suggesting coexistence of the
phases similarly as in the case of partial polymerization.}
\label{Figure7}
\end{figure}

The polymeric phase remains stable up to the depolymerization
temperature, $T_{DP}= 520$ K.  Above this temperature, both the
$\Delta H_{pp}$ and the $g$-factor recover their characteristic
values for the ferromagnetic phase (Fig.~\ref{Figure7}).  The
depolymerization process is an irreversible transition. At the
depolymerization temperature, both polymeric and ferromagnetic
ESR features are present, thus pointing to the coexistence of the
two phases. It also suggests that this phase transition does not
seem to be an abrupt one. As in the case of ferromagnetism, the
polymerization process might be influenced by percolation
mechanism, as the samples of apparently lesser quality exhibit a
bit lower depolymerization temperature.

At low temperatures, the depolymerized crystals reveal the same
temperature dependences of the ESR parameters as those observed
for the ferromagnetic crystals. The $g$-factor shifts remarkably
below 15 K, which indicates that the depolymerized crystal
undergoes the ferromagnetic phase transition.

\section{Discussion}

In the intermediate pressure range, below $P_C = 7$ kbar, the
TDAE-C$_{60}$ monomer partially polymerizes on cooling.  Both the
polymerization temperature ($T_P$) and the relative ratio of the
polymeric to the monomeric fractions depend on pressure
(Fig.~\ref{Figure4}). Preliminary analysis\cite{17} has suggested
that the polymer would have a linear [2+2] cycloadduct bonding
structure with a similar intra-chain ball distance as in the case
of Rb$_1$C$_{60}$ (Ref. \cite{20}). The most prominent difference
between the systems is a much larger inter-chain distance in the
case of TDAE-C$_{60}$, which is due to the large size and
anisotropic (steric) properties of the TDAE interstitials.

Rich phase diagrams have already been reported for several
C$_{60}$-related compounds. In particular, interesting phase
transitions were found for Rb$_1$C$_{60}$. Depending on cooling
rate and quenching, it can form either monomeric,
dimeric\cite{21} or polymeric phases\cite{20}\cite{22}.
Nevertheless, the coexistence of the two phases has never been
observed for Rb$_1$C$_{60}$. In contrast, the coexistence of two
phases at low temperatures was observed in the case of
single-bonded linear polymers of Na$_2$RbC$_{60}$
system\cite{23}. In that case, the partial polymerization can be
explained by steric effects and by disorder resulting from
different possible directions of the bond formation. In contrast,
TDAE-C$_{60}$ is a strongly anisotropic structure with the only
one possible direction of the bond formation (c-axis). Hence, the
partial polymerization is probably governed by the orientation
disorder of JTD-C$_{60}$ molecules. The polymerization
temperature, $T_P$, is much lower than the freezing temperature
of the C$_{60}$ molecule rotation for a given pressure. This
suggests that the freezing of molecular orientations is not
sufficient for initiating the polymerization process and there
exists a kinetic barrier for this phase transition.  It seems to
be natural that different kinetic barriers can characterize
various relative orientations of the JTD-C$_{60}$ molecules.  As
the remaining monomeric phase still shows a pronounced
ferromagnetic order, most probably, the FM molecular
configuration has the highest energetic barrier.

It follows from our measurements that the depolymerization of
TDAE-C$_{60}$ occurs at the higher temperature than in the case
of one-dimensional A$_1$C$_{60}$ polymers (A= K, Rb, Cs), where,
depending on the compound, $T_{DP}$ varies in the range of $300 -
400$ K (Ref. \cite{4}).  It seems that $T_{DP}$ of TDAE-C$_{60}$
is rather comparable to that found for the two-dimensional
polymer Na$_4$C$_{60}$ ($T_{DP} \approx 500$ K)\cite{24}.
Therefore, the polymeric chains of TDAE-C$_{60}$ seem to form
much more stable structures than the double-bonded polymeric
chains of the Rb$_1$C$_{60}$ system.  If the intra-chain bounds
of both TDAE-C$_{60}$ and Rb$_1$C$_{60}$ were of the same nature
(isostructural), one would expect for them similar temperature
stability.  The observed discrepancy in the temperature stability
can be ascribed to a potential structural difference in the two
polymeric structures. Alternatively, the previously not
investigated effects of dopant molecule and the inter-chain
coupling can be important for the polymeric chain stability. A
precise structural analysis is needed to answer these questions.

Below room temperature ($T < 260$ K), the ESR spin susceptibility
of the polymerized $\beta-$phase follows a simple Curie law with
one spin $S=1/2$ per chemical formula. This indicates that spins
probed by the ESR experiment are localized. Moreover, in this
temperature range, the ESR linewidth for the polymerized
$\beta-$phase is very narrow ($0.5$ G) and almost temperature
independent, which suggests strong exchange interactions between
the spins.  The theoretical band calculations, however, predict
that in the absence of electron correlations, the isolated
single-charged, double-bonded linear polymer should be metallic
with a half-filled band\cite{25}. Also, the same property holds
well for single-bonded linear polymers encountered in
Na$_2$AC$_{60}$ systems\cite{26}. Indeed, all the other charged
linear polymers of the C$_{60}$ compounds, discovered to date,
are metallic in a wide temperature range, with possible
ground-state instabilities, such as spin-density wave\cite{27}.
Therefore, the TDAE-C$_{60}$ polymeric phase seems to be unique.

The inter-chain coupling is very important in alkali fullerides
A$_1$C$_{60}$, influencing their dimensionality. However,
according to Erwin et al. \cite{25}, a direct inter-chain
coupling interaction can be neglected in TDAE-C$_{60}$ based on
the large inter-chain separation found in these systems.

Assuming a simple model, one would expect the TDAE-C$_{60}$
polymer should be a strongly anisotropic metal. The effective
strong localization observed in the $\beta-$phase of
TDAE-C$_{60}$ might suggest that its actual polymer topology
essentially differs from that of Rb$_1$C$_{60}$. Alternatively,
the localization effect might originate from the possible
enhancement of the effective Coulomb repulsion at C$_{60}$ sites,
due to the influence of the TDEA$^+$ radicals. In such a
Mott-Hubbard localization scheme, one would expect a
non-vanishing Weiss constant.  In contrast, the presence of the
Weiss constant has not been detected in the low-temperature
$\beta-$phase. Mizoguchi et al.\cite{17} suggested that the
absence of the Weiss parameter could be explained by
re-activation of the positive exchange coupling with the spins of
TDAE$^+$ radicals. This, in turn, would fortuitously cancel the
negative inter-C$_{60}$ spin coupling leading to a diminishing
Weiss constant. Nevertheless, in this work, we did not observe
the complete recovery of the TDAE$^+$ spins in the polymeric
phase at the room temperature, as claimed by Mizoguchi et al.
Indeed, the appearance of TDAE-spins at elevated temperatures is
accompanied with development of the antiferromagnetic Weiss
constant.

Assuming that polymeric chains in the $\beta$-TDAE-C$_{60}$ have
rather small sizes and are disordered, one can apply a model of
the random-exchange AFM Heisenberg 1D chains. In this case, the
Weiss temperature would be absent, whereas the ESR susceptibility
should be proportional to $T^{-a}$, where $\alpha \approx 0.7 -
0.8$. Fitting this model to our data yields $\alpha = 0.96 \pm
0.1$. Clearly, this result does not support the above-mentioned
model. For right now, a plausible reason for non-appearance of
the Weiss temperature in the polymer phase remains unclear.

As can be seen in Fig.~\ref{Figure6}a, in the polymeric
$\beta-$phase, a complete recovery of the TDAE$^+$-related spins
is not observed at ambient temperature. The recovery of the spins
occurs rather at higher temperatures, with full development at
320 K. The dynamics of the TDAE$^+$-related spins is probably
connected to small movements of the TDAE$^+$ molecules leading to
the dimerization shift\cite{13}.\\

\section{Conclusion}

In conclusion, we have investigated the polymerization mechanism
in TDAE-C$_{60}$ ferromagnetic system and the physical properties
of the polymeric $\beta$-phase. The complete polymerization at
room temperature is observed at the pressure of 7 kbar. At the
same pressure, the ferromagnetic transition is suppressed. Partial
polymerization is observed at lower pressures and temperatures,
and the stability of the polymeric phase was established in the
$P - T$ parameter space. The high depolymerization temperature
suggests that the polymeric chains are much more stable than in
the case of double-bonded linear polymers of the Rb$_1$C$_{60}$
systems. To deduce the exact structural properties of the
polymer, an additional high-resolution X-ray diffraction study is
needed.

Moreover, the observed strong localization of spins in the
polymeric TDAE-C$_{60}$ is in contradiction with conclusions from
a simple theoretical reasoning and it is not comparable to any
other charged linear polymer of the C$_{60}$. Above 320 K, the
previously silent spins on the TDAE-C$_{60}$ are revealed. The
observed decoupling of the TDAE$^+$ spins is accompanied by the
appearance of the Weiss constant.

In summary, we conclude that the polymerization phenomenon in
TDAE-C$_{60}$ is very important to better understanding the
ferromagnetic properties of the system. However, it can also shed
new light towards our general understanding of the ground state
electronic properties of the linear fullerides polymers.

\begin{acknowledgments}
This work has been supported in part by a Grant-in-Aid for
Scientific research on the priority area ``Fullerenes and
Nanotubes'', from the Japanese Ministry of Education and Culture
(T.K., M.F., and K.O.) and by the Polish KBN Grant
\#2-PO3B-090-19 (A. S.). The support of the Swiss National
Scientific Foundation is also greatly acknowledged (S. G. and L.
F.).
\end{acknowledgments}

\newpage
\end{document}